**World Scientific**
www.worldscientific.com

# FOXSI-2: Upgrades of the Focusing Optics X-ray Solar Imager for its Second Flight

Steven Christe[*,‡‡], Lindsay Glesener[†,‡], Camilo Buitrago-Casas[†], Shin-Nosuke Ishikawa[¶],
Brian Ramsey[§], Mikhail Gubarev[§], Kiranmayee Kilaru[§], Jeffery J. Kolodziejczak[§],
Shin Watanabe[¶,‖], Tadayuki Takahashi[¶,‖], Hiroyasu Tajima[**], Paul Turin[†],
Van Shourt[†], Natalie Foster[†] and Sam Krucker[†,††]

[*]NASA Goddard Space Flight Center
Greenbelt, MD 20771, USA

[†]Space Sciences Laboratory
University of California at Berkeley, Berkeley, CA, USA

[‡]University of Minnesota-Twin Cities, Minneapolis, MN, USA

[§]NASA Marshall Space Flight Center, Huntsville, AL, USA

[¶]Institute of Space and Astronautical Science (ISAS)/JAXA, Sagamihara, Japan

[‖]Department of Physics, University of Tokyo, Tokyo, Japan

[**]Institute for Space-Earth Environmental Research
Nagoya University, Nagoya, Japan

[††]University of Applied Sciences and Arts Northwestern Switzerland
Windisch, Switzerland
[‡‡]steven.christe@nasa.gov

The Focusing Optics X-ray Solar Imager (FOXSI) sounding rocket payload flew for the second time on 2014 December 11. To enable direct Hard X-Ray (HXR) imaging spectroscopy, FOXSI makes use of grazing-incidence replicated focusing optics combined with fine-pitch solid-state detectors. FOXSI's first flight provided the first HXR focused images of the Sun. For FOXSI's second flight several updates were made to the instrument including updating the optics and detectors as well as adding a new Solar Aspect and Alignment System (SAAS). This paper provides an overview of these updates as well as a discussion of their measured performance.

## 1. Background

The processes of particle acceleration and impulsive energy release in magnetized plasmas is one of the outstanding mysteries in heliophysics and astrophysics. It is generally recognized that the energy source must be the magnetic field but the process through which particles are accelerated to high energies is still not well understood. Hard X-ray (HXR) observations are a powerful diagnostic tool that provides quantitative information such as the energy and location of non-thermal accelerated electrons ($> 10\,\mathrm{keV}$).

Current observations have confirmed that the acceleration region for flares is located high in the tenuous solar corona where the density is relatively low. This fact means that the expected HXR bremsstrahlung emission is faint. Because of this, observations of these kinds of sources are relatively rare (see Krucker *et al.*, 2008). In addition, these observations are also hard to make due to the fact that they are frequently accompanied by much brighter X-ray emission from the chromosphere. Past hard X-ray observations of the Sun (e.g. RHESSI, Lin *et al.*, 2002; Acton *et al.*, 1992) have relied on indirect imaging techniques (see Hurford *et al.*, 2002) which have difficulty with providing high sensitivity combined with high imaging dynamic range.

Table 1.   FOXSI mission overview.

| Parameter | Value |
| --- | --- |
| NSROC ID | 36.295 |
| Launch site | White Sands Missile Range |
| Rocket type | Terrier Black Brant IX |
| Launch time (UT) | 2014-12-11 19:11:01 |
| Apogee | 339 km |
| Observation time (UT) | 19:12:51 to 19:19:25 |

The Focusing Optics X-ray Solar Imager (FOXSI) is a sounding rocket payload part of a program to develop and apply new HXR grazing-incidence focusing optics combined with solid-state pixel detectors to solar observations (also see Christe *et al.*, 2013; Gaskin *et al.*, 2013 for an overview of the HEROES balloon payload part of the same program). FOXSI launched from the White Sands Missile Range for the second time on 2014 December 11 at 19:11 UT and reached an apogee of 339 km. For a more in-depth overview of FOXSI, its first flight, and scientific results see Krucker *et al.* (2009, 2013, 2014) and Ishikawa *et al.* (2014). An overview of the mission parameters for FOXSI's second flight are shown in Table 1.

This paper will provide a technical overview of the FOXSI payload focusing on the systems that were upgraded or added for its second flight. The scientific observations of the flight will be reported in a future paper.

## 2.   FOXSI Payload Overview

The FOXSI payload is composed of seven similar HXR grazing-incidence telescope modules with a focal length of 2 m. Each telescope module is composed of a number of co-aligned and nested mirror shells. Each telescope module focuses X-rays onto a dedicated solid-state double-sided strip detector. An overview of the instrument can be seen in Table 2.

A rendering of the overall payload inside the rocket skins is shown in Fig. 1. The FOXSI payload is cantilevered to the rocket skin at the optics plane. The seven optics modules are mounted to the optics plane at their center of mass. The optics plane also accommodates the two NSROC-provided Solar Pointing Attitude Rocket Control System (SPARCS VII) sensors (Lockheed Intermediate Sun Sensor, LISS, & Miniature Acquisition Sun Sensor, MASS) as well as the front lens assembly and baffle for the SAAS. The detector plane holds each of the seven detectors and provides passive cooling for the detectors during the flight. The SAAS camera and camera lens assembly are also mounted to the detector plane in addition to a quad photocell used for ground alignment which lies behind a pinhole in the optics plane.

Table 2.   FOXSI-2 instrument overview.

| Parameter | Value |
| --- | --- |
| Focal length | 2 m |
| Optics configuration | Wolter I |
| Number of optics modules | 7 |
| Number of mirror shells | 7 (5), 10 (2) |
| Detector type | Double-sided strip |
| Detector material | Si (5), CdTe (2) |
| Strip pitch | 75 $\mu$m  (i), 60 $\mu$m (CdTe) |
| Angular resolution | |
|   Optics (FWHM) | 5 arcsec (optics) |
|   Plate Scale | 7.7 arcsec  (Si) |
| | 6.2 arcsec (CdTe) |
|   Overall[a] | 9.2 arcsec (Si), 8.0 arcsec (CdTe) |
| Strip number | 128 × 128 |
| Detector dimensions | 9.6 mm$^2$ × 9.6 mm$^2$ (Si), |
| | 7.68 mm$^2$ × 7.68 mm$^2$ (CdTe) |
| Field of view | 16.5 arcmin × 16.5 arcmin (Si) |
| | 13.2 arcmin × 13.2 arcmin (CdTe) |
| Energy range | ~4 to 15 keV |
| Effective area | 115 cm$^2$ up to 10 keV |

[a]Optics FWHM added in quadrature with detector strip pitch.

Most of the detectors (5 out of 7) are made of silicon (Si) and have a strip pitch of 75 microns (equivalent to 7.7 arcsec over 2 m). For the second flight of FOXSI, two detectors were replaced with cadmium telluride (CdTe) versions of the same detectors with a strip pitch of 60 microns (equivalent to 6.2 arcsec over 2 m). The angular resolution of the optics (defined here as the Full Width Half Maximum, FWHM, of the Point Spread Function, PSF) were measured to be ~5 arcsec (see Krucker *et al.*, 2014). Two of the optics modules were updated from seven nested shells to 10 nested shells in order to increase the effective area. Measurements of the angular resolution confirmed that the additional shells did not measureably degrade the resolution of the 10-shell optics. The overall angular resolution of the telescope modules with the combined optics and detector resolution is 9.2 arcsec (Si) and 8.0 arcsec (CdTe) where this value is calculated by adding the FWHM and plate scale in quadrature.

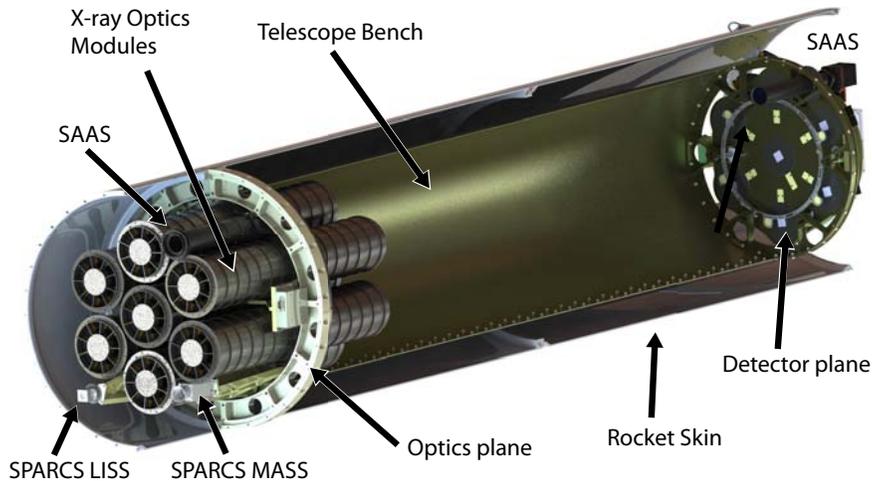

Fig. 1.   A rendering of the FOXSI payload. The payload is cantilevered to the rocket skin at the optics plane. The payload consists of seven telescope modules each consisting of an optics module at the optics plane and a corresponding detector at the detector plane. The detector plane is connected to the optics plane through an Aluminum telescope bench. The SPARCS sun-pointing system which consists of the LISS and MASS sensors are mounted to the optics plane on towers to provide them an obstructed view of the Sun. The Solar Aspect and Alignment System (SAAS) consists of a lens coupled to a baffle mounted to the optics plane as well as a camera with a compound lens system mounted to the detector plane.

## 3.  FOXSI-2 Updates

### 3.1.  *X-ray optics*

The FOXSI mirror shells are Wolter type 1 geometry (Werner, 1977) and are monolithic structures containing both the parabolic first mirror and the hyperbolic second mirror segments. These mirrors were manufactured using an electroformed nickel replication process developed at the NASA Marshall Space Flight Center (MSFC) (see O'Dell *et al.*, 2015 for a review of the optics program) whereby the shells are electro-deposited onto superpolished mandrels. Once formed, the shells are released from the mandrel by cooling in a chilled-water bath. This process is relatively cost-effective compared to conventional polishing and figuring of individual heavy mirrors and produces shells with inherently good angular resolution. In order to build up collecting area, individual mirror shells are nested together to produce an optic module. A key advantage of this technique is that multiple mirror shells can be produced using a single mandrel which means that the incremental cost of producing an identical optic module is small. Similar optics produced by the same group have been flown on the High Energy Replicated Optics (HERO) (Ramsey *et al.*, 2002), and High Energy Replicated Optics to Explore the Sun (HEROES) (Christe *et al.*, 2013; Gaskin *et al.*, 2013) balloon payloads and will soon be flown as part of the ART-XC Instrument (Pavlinsky *et al.*, 2011) on board the Spectrum Roentgen Gamma (SRG) mission, an X-ray astrophysical observatory, developed by Russia in collaboration with Germany.

One of the principle attractions of this technique for solar observations is that it provides superior angular resolution, an important consideration because the Sun is close enough that some important spatial scales can frequently be resolved unlike astrophysical sources. In the HXR range, angular scales are frequently observed and resolved with sizes above a few arcsec (Warmuth & Mann, 2013a,b; Kontar *et al.*, 2010) though smaller scales may also be present. The resolution of these optics is made possible by the fundamental stability of the monolithic optic configuration. The shells have an inherent resolution matching that of the polished mandrel. Additionally, nested optics require that each optic be precisely aligned to each other otherwise the overall resolution of an optic module will be degraded compared to the resolution of individual optic shells. Other techniques such as segmented optics must align many more elements. For example, NuSTAR (Harrison *et al.*, 2013) which uses slumped glass segmented optics mirrors uses sextants (60° sectors) or twelvetants (30° sectors) for both the primary and secondary mirrors. That means that to build up a single optic shell 12 to 24 individual mirror segments must be aligned to each other. The NuSTAR measured resolution is

Table 3. Optics configuration overview for a single optic module. Each shell has a focal length of 2 m, a thickness of 0.25 mm and a segment length of 30 cm. Each is coated with 30 nm of Iridium. The FOXSI-2 upgrade added three inner shells (#8, #9, #10) to two of the optics (X-0, X-2).

| Shell # | Outer diameter (mm) | Inter diameter (mm) | Inner diameter (mm) | Graze angle (°) | Mass (kg) | Geometric area (cm$^2$) |
|---|---|---|---|---|---|---|
| 1 | 106.9 | 103.0 | 91.2 | 0.37 | 0.43 | 6.37 |
| 2 | 101.7 | 98.0 | 86.7 | 0.35 | 0.41 | 5.76 |
| 3 | 96.7 | 93.2 | 82.5 | 0.33 | 0.39 | 5.21 |
| 4 | 91.9 | 88.6 | 78.4 | 0.32 | 0.37 | 4.71 |
| 5 | 87.3 | 84.2 | 74.5 | 0.30 | 0.35 | 4.25 |
| 6 | 83.0 | 80.0 | 70.8 | 0.29 | 0.34 | 3.84 |
| 7 | 78.8 | 76.0 | 67.2 | 0.27 | 0.32 | 3.47 |
| 8* | 74.5 | 71.8 | 63.6 | 0.26 | 0.30 | 3.09 |
| 9* | 70.2 | 67.6 | 59.9 | 0.24 | 0.28 | 2.74 |
| 10* | 65.8 | 63.4 | 56.1 | 0.23 | 0.27 | 2.41 |
| | | | | Total | 2.61, 3.46* | 33.6, 41.9* |

*Note*: Enhanced configuration for optic X-0 and X-2.

18 arcsec FWHM and ∼58 arcsec Half-Power Diameter (HPD) (Westergaard *et al.*, 2012; Madsen *et al.*, 2015) while FOXSI has values of 5 arcsec and 25 arcsec, respectively. This is essentially the same as an individual mirror shell, meaning that the alignment process does not degrade the resolution.

FOXSI consists of seven telescope modules each with 7 or 10 shells. These modules are referred to as X-0 through X-6. For FOXSI-2, two of the telescope modules (X-0, X-2) were upgraded to include 10 shells. The shell configuration is described in Table 3. The average spacing between the shells is about 2 mm. The total weight of the shells (neglecting mounting structure) is about 3.5 kg. The graze angles range from 0.37° (for the largest shell) to 0.23 (for the inner shell). The shells numbered 1 to 7 are the original shells. These shells provide a geometric area of 33.6 cm$^2$. The FOXSI-2 upgrade added 3 inner shells (#8, #9, #10) providing an additional 8.3 cm$^2$ of geometric area (a 25% increase) at an added weight of 0.85 kg.

The method through which these shells are mounted and aligned is described in Gubarev *et al.* (2009).

## 3.2. *Effective area*

The performance of the FOXSI optics was measured at the Stray Light Test Facility at MSFC. This facility consists of a 100 m evacuated beam line. A Trufocus 50 kV microfocus X-ray generator with a titanium target is placed at one end of the facility and is used to illuminate a FOXSI optics module inside an evacuated chamber. To measure the effective area, an Amptek XR-100T cadmium-zinc-telluride (CdZnTe) detector is placed directly behind the evacuated chamber behind a small air gap (∼ 1 cm). Above 5 keV, absorption by this column of air is ≲ 5%. The position of the detector is adjusted to find the focus. The effective area is determined by measuring the focused flux and dividing by a measurement of the flux without the optic. At finite source distances the measured effective area is slightly larger (∼5%) than for a source at infinity, because more X-rays can intersect both the first parabolic and the second hyperbolic surfaces. This effect has been corrected for in the results shown here.

A plot of the measured total effective area provided by all optic modules is in Fig. 2. The measured effective area for a representative 7-shell optic (X-4) and 10-shell (X-0) optic are also shown for comparison. The effective area is relatively constant from 5 keV to ∼10 keV, with a total effective area of about 115 cm$^2$. The effective area then decreases due to decreasing reflectivity limited by the 2-m focal length of the FOXSI optics. At 14 keV, the effective area is still ∼20 cm$^2$.

The off-axis performance was also investigated. Changes to the throughput as a function of energy are expected as the graze angle changes for off-axis sources. The optics were mounted on tip-tilt stages so that the source could effectively be placed at any off-axis angle. In this study, we measured the performance up to 9 arcmin. This value is similar to the field of view provided by the detectors (Si 16.5 × 16.5 arcmin edge to edge, CdTe 13.2× 13.2 arcmin).

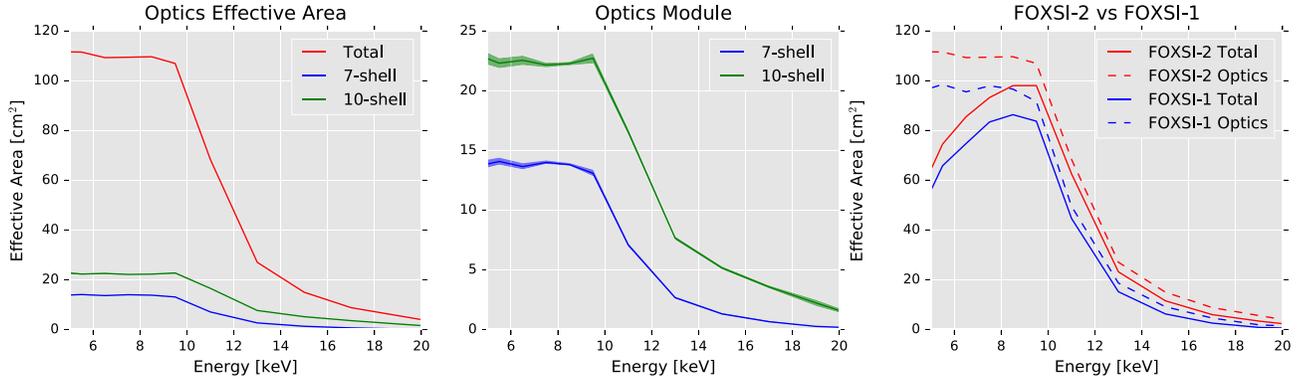

Fig. 2. (*Left*) A plot of the total measured FOXSI on-axis effective area provided by the sum total of all of the optics modules. The on-axis effective area provided by a representative 7 shell optic, and 10 shell optic are also shown for comparison. The shaded area around the lines represent the errors in the measurements. The total on-axis effective area is about 115 cm$^2$ and is fairly constant up to around 10 keV where it begins to decrease due to decreasing reflectivity provided by the mirrors at these energies. The 7-shell optics provide ~14 cm$^2$ each while the 10-shell optics provide ~23 cm$^2$ each. (*Middle*) A comparison of the effective area of 7 and 10 shell optic. (*Right*) A comparison between the effective area provided by FOXSI-1 and FOXSI-2 including detector efficiency which reduces the area at high energies as well as material in the optical path which reduces the area at low energies. For comparison the optics area is also shown as dashed lines.

As the source moves off-axis the effective area decreases due to the increase in the angle of the incoming rays with the optics and the decrease in the projected geometric area of the mirror surfaces. For large enough angles outer mirror shells begin to obstruct inner shells, further decreasing the throughput. Figure 3 shows the effective area as a function of energy for various off-axis angles in both positive and negative tip and pan directions for one of the modules. Effective areas for positive and negative off-axis angles agree with each other to within the errors, suggesting that the physical properties of the optics modules are symmetric. It is found that the field of view (FWHM) of the optics module is greater than 9 arcmin.

### 3.3. Detectors

#### 3.3.1. Requirements

Future solar-dedicated X-ray instruments require detectors with fine spatial and energy resolutions, and low electronic noise, power consumption and backgrounds. An energy resolution better than 1 keV is needed to differentiate thermal and non-thermal bremsstrahlung contributions. To meet these requirements, *FOXSI* uses thin double-sided semiconductor strip detectors with small strips (60–75 $\mu$m) and high sensitivity in the 4–15 keV energy range. These detectors were provided by the Institute of Space and Aeronautical Sciences (JAXA/ISAS) in Japan, as part of the development for the Soft-gamma ray detector (SGD) that will fly onboard the *Astro-H* spacecraft (Takahashi *et al.*, 2014; Sato *et al.*, 2014).

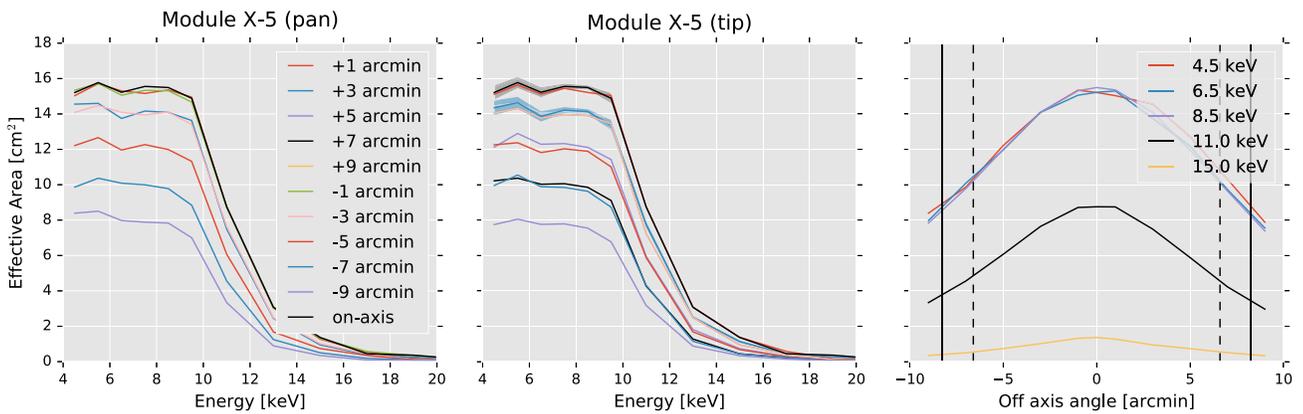

Fig. 3. (Color online) (*Left*) The measured effective area as a function of energy of a single FOXSI optics module (X-5) for different off-axis angles. The left plot shows off-axis angles in the horizontal direction (pan) while the center plot shows them for the vertical direction (tip). The colored areas around the lines show the error in the measurements. The measurements of the effective area for the absolute value of each off-axis angle agree with each other within the measurement errors, showing that the response is highly symmetric about the telescope optical axis. (*Right*) The effective area as a function of off-axis angle for the same optic. The effective area varies smoothly from the on-axis position. The vertical unbroken and dashed gray lines represent the edges of the Si and CdTe detectors, respectively. The effective area is seen to decrease to ~50% at the edge of the Si detector and to ~60% at the edge of the CdTe detector. Larger decreases are expected at the corners of the detectors.

The original angular resolution requirement for the detectors was to have a smaller strip pitch (pixel size) than the optics FWHM, which was expected to be ∼12 arcsec though in actuality the optics exceeded expectations. To meet this requirement, *FOXSI*-1 used silicon (Si) strip detectors composed of two perpendicular sets of 75 $\mu$m pitch strips (see Fig. 4) in a 128 by 128 array, corresponding to a 7.7 arcsec resolution at a focal length of 2 m. For *FOXSI*-2, detector upgrades were motivated by two desires: (1) to develop and test detectors for future solar observations that are efficient up to ∼50 keV; this excludes Si; and (2) to achieve better angular resolution in order to oversample the optics PSF. A 60 $\mu$m pitch CdTe detector, also composed of double-sided strips, was selected to meet these requirements. See Ishikawa *et al.* (2010, 2011) for detailed information on the *FOXSI*-1 and *FOXSI*-2 detectors, and Glesener (2012) for a full description of the *FOXSI* detector/readout system.

### 3.3.2. *Operation and readout*

The Si and CdTe detectors are read out using customized low-power, low-noise ASICs; these were designed as a joint collaboration between JAXA/ISAS, Stanford and Gamma-Medica Ideas (Watanabe *et al.*, 2009, 2014). The ASICs have 64 input channels, and two ASICs are used to read out each side of the detector. An iteration designed especially for *FOXSI*'s low-energy needs (the VATA451) and an iteration designed for Astro-H's Hard X-ray Imager (HXI) (the VATA450) were used for the Si and CdTe detectors, respectively. The ASIC includes a charge-sensitive amplifier, fast and slow shapers for triggering and measurement, and a Wilkinson-style ADC for each channel, enabling the measurement and digitization of all channels in parallel. The ASICs have a power consumption of 1 mW per channel (∼0.26 mW per detector). The telemetry stream allows 500 frames per second. The *FOXSI* payload includes no onboard processor and no memory storage for multiple events, so no more than one event per detector can be recorded in each 2 ms telemetry frame. This is sufficient for measurement of the quiet Sun and quiescent active regions in the 4–20 keV with *FOXSI*'s effective area. To ensure low-noise performance, the detectors were maintained below −15° during the observing portion of the flight.

The detectors are operated at a bias voltage of 200 V. The voltage was ramped from 0 starting 30 s after launch at a ramp speed of 5 V s$^{-1}$ so that the detector current had time to stabilize before observations began.

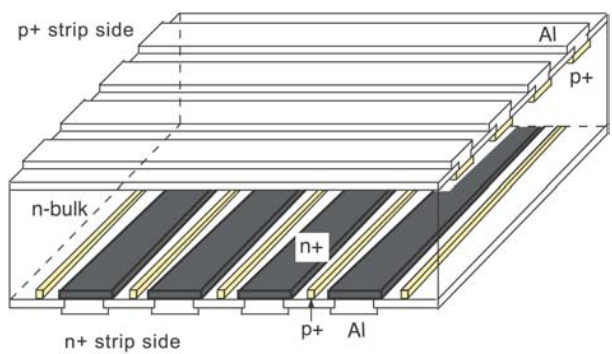

Fig. 4. Diagram of a double-sided strip detector. Holes and electrons are collected at strips on the p- and n-sides that are oriented orthogonally to each other so that a two-dimensional image can be obtained. Each strip acts as an individual p-n junction. Figure is taken from Takeda *et al.* (2008).

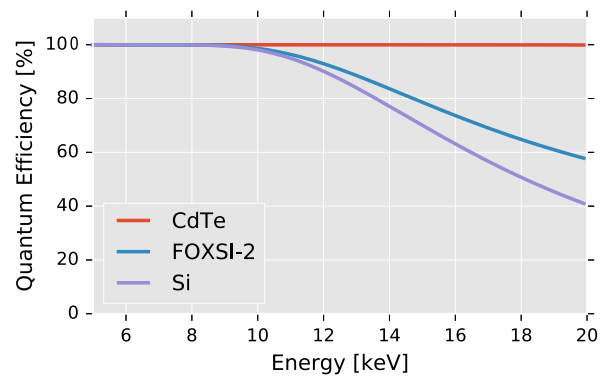

Fig. 5. The quantum efficiency of the 0.5 mm thick FOXSI-2 detectors. Silicon (Si) efficiency falls to ∼40% at 20 keV while the CdTe efficiency remains almost unity throughout the FOXSI energy range. The average quantum efficiency for FOXSI-2, which contains 2 CdTe and 5 Si detectors, is also plotted.

### 3.4. *The SAAS*

A key challenge for X-ray telescopes is the alignment between the X-ray optics and the aspect system used to provide solar pointing control and knowledge. For solar-pointed sounding rocket payloads, the primary solar aspect system is the NASA-provided Solar Pointing Attitude Rocket Control System (SPARCS VII). SPARCS is a three-axis solar pointing control system comprised of a coarse (CSS), intermediate (MASS), and fine Sun sensor (LISS). It makes use of a pneumatic system and ring laser gyros to control attitude. SPARCS can achieve pointing accuracy of 30 arcsec in pitch and yaw and 1.2° in roll. Pointing control stability of 0.12 arcsec (at the $2\sigma$ level) was seen in the FOXSI-1 flight.

In order to improve upon a past alignment procedure (see Krucker *et al.*, 2013), a new SAAS was developed. The SAAS consists of a co-boresighted optical telescope which is physically coupled to both the optics and detector plane. The primary purpose of the SAAS is to provide a measurement of the alignment between the SPARCS LISS and the FOXSI X-ray optics. To achieve this it must

- provide a reference sensor for alignment during FOXSI X-ray alignment and
- provide a reference image of the Sun at the same time as the SPARCS ground alignment.

An optional third purpose is to provide real-time pointing feedback during the flight. The system requirements are

- the SAAS shall provide a field of view of no less than $2° \times 2°$.

- the SAAS shall provide an alignment between the X-ray optics and the SPARCS LISS to better than 5 arcmin.

An additional desire was to keep the system as simple as possible. A single lens system was deemed unfeasible because the field of view requirement would require at least 70 cm of available space in the detector plane which was not available.

The optical solution for the SAAS consists of three lenses, the primary lens or solar lens ($f = 1500$ mm, LPX-25.0-778.1-C from IDEX) is mounted on the optics flange, and two additional lenses, camera lens 1 ($f = 200$ cm, 011-2670 from Optosigma, now SLB-40-200P) and camera lens 2 ($f = 30$ cm, LB1757 from Thorlabs) are mounted to the detector flange (see Fig. 6). In order to reduce cost all lenses are off-the-shelf and the lens material is borosilicate glass Schott (BK7).

A set of filters were placed in front of the solar lens. This includes an IR cut-off filter (Edmund Optics 53-710) for heat rejection, a 632 nm hard coated bandpass filter (Edmund Optics 65-166), and three neutral density filters (ND 4, ND 0.6, ND 0.3) which were only used during solar observations. The final filter (ND 0.3) was only used during flight since the Sun's intensity is about twice as strong outside the Earth's atmosphere. The bandpass filter was chosen to provide a sharp solar limb and reduce chromatic aberration (Henneck *et al.*, 1999; Fivian *et al.*, 2002).

The CCD camera used was the Imperx Bobcat IGV-B1310. It is a fully programmable CCD camera built around the SONY ICX 445 3.75 micron

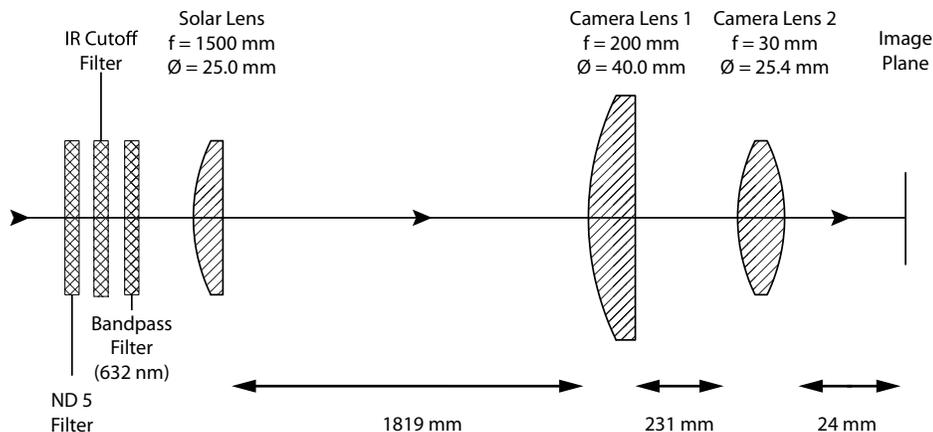

Fig. 6. The optical design for the SAAS. The system is composed of three lenses, of which are plano-convex. The solar lens is mounted on the optics bench while the two camera lenses are mounted at on detector bench 2 m away. A set of three filters are used to reduce the intensity (ND filter), reduce heat flux (IR Cutoff filter) and reduce chromatic aberration (bandpass filter). The ND filter is actually composed of three different filters (ND 4, ND 0.6, ND 0.3).

Interline Transfer CCD image sensor (1/3 in optical format). It provides an image resolution of 1296 × 966 (3.75 micron pitch pixels) and up to 39 frames/s at full resolution. A ADLINK PC/104 computer (Cool RoadRunner-945GSE) was used to control and readout the camera. Due to the limited storage capacity of the system no images could be stored during flight.

In order to provide real-time pointing feedback during the flight a Advanced Micro Peripherals nanoVTV board was used to convert the output VGA signal from the single board computer to an NTSC TV signal. This signal is telemetered to the ground in real-time during flight though with lossy compression applied. Because of this SAAS flight data are not of any scientific value. A future version will include on-board storage in order to save full resolution data.

The system was designed and tested in the Zemax software. Simulations of the optical performance suggested that the best achievable resolution is about two pixels (∼6 arcsec). In order to achieve this resolution assuming mechanical tolerances required the following adjustments be included as part of the system.

- The airspace between the solar lens and camera lens 1 shall be adjustable to ±1 mm in 0.1 mm steps.
- The airspace between camera lens 2 and the detector shall be adjustable to ±1 mm in 0.1 mm steps.
- The X/Y center of the Solar lens shall be adjustable to ±0.5 mm in 0.1 mm steps.

Due to limitations on the physical space available for the camera lenses, some vignetting was expected. The system was first integrated in the field and limited time was available to focus the system. The system was focused by observing the Organ mountain range about 10 miles away from the White Sands Missile Range Vehicle Assembly Building where the SAAS was integrated. Unfortunately seeing was significantly degraded by warm air currents at the time of the observation which limited the ability to focus the system. An analysis of the angular resolution of the system is discussed in Sec. 3.6.

### 3.5. *X-ray alignment procedure and results*

The alignment is measured through the use of a precision machined alignment fixture which precisely located two laser sources as well as an X-ray generator (see Fig. 7). This fixture is placed 20 m away from the front of the FOXSI payload. Larger distances would provide more accurate alignment since any positional errors translate to angular errors that decrease linearly with distance but X-ray absorption by the air increases exponentially and therefore becomes prohibitively large at greater distance. At 20 m a 1 mm misalignment corresponds to an error of 10.3 arcsec. The positions of the mounting points on the alignment fixture are known to machine tolerances or 25 microns (2.5 arcsec at 20 m). The alignment fixture is placed in position using a quad photo-cell which is directly behind a pinhole in the optics plane. This reference sensor provides a quick alignment so that it is known that the alignment fixture is in the correct position to be seen by both the SAAS and the X-ray optics. For more information about this alignment procedure see Krucker *et al.* (2013).

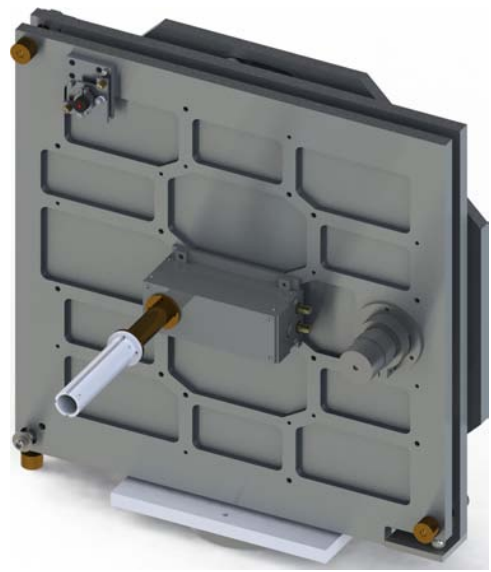

Fig. 7. A render of the alignment fixture. The fixture holds an X-ray source as well as two laser sources. In this render, the X-ray source would illuminate the center optic but can be moved to any other optic position. A laser source (upper right) is used to illuminate the SAAS and another laser source (right) is used to illuminate a quad photo-cell and is used for the initial placement of the fixture 20 m away from the FOXSI optics. Each laser source sits behind a small pinhole (∼50 micron).

To perform a measurement of the alignment between the X-ray optics and the SAAS, a laser mounted on the alignment stand is shown through a small pinhole (∼50 micron). The SAAS then images the location of that laser spot. An image of this can be seen in Fig. 10. The laser spot is a classic Fresnel diffraction pattern. In order to determine the laser spot center a fit is made to the outer ring of the diffraction pattern. The fit consists of a circularly symmetric Gaussian with a linearly varying amplitude to correct for uneven illumination of the aperture. This fit determines the center to within 2 pixels (or ∼6 arcsec). A more accurate model of the diffraction pattern would likely yield a more accurate fit. This fit determines the calibrated center of the SAAS CCD.

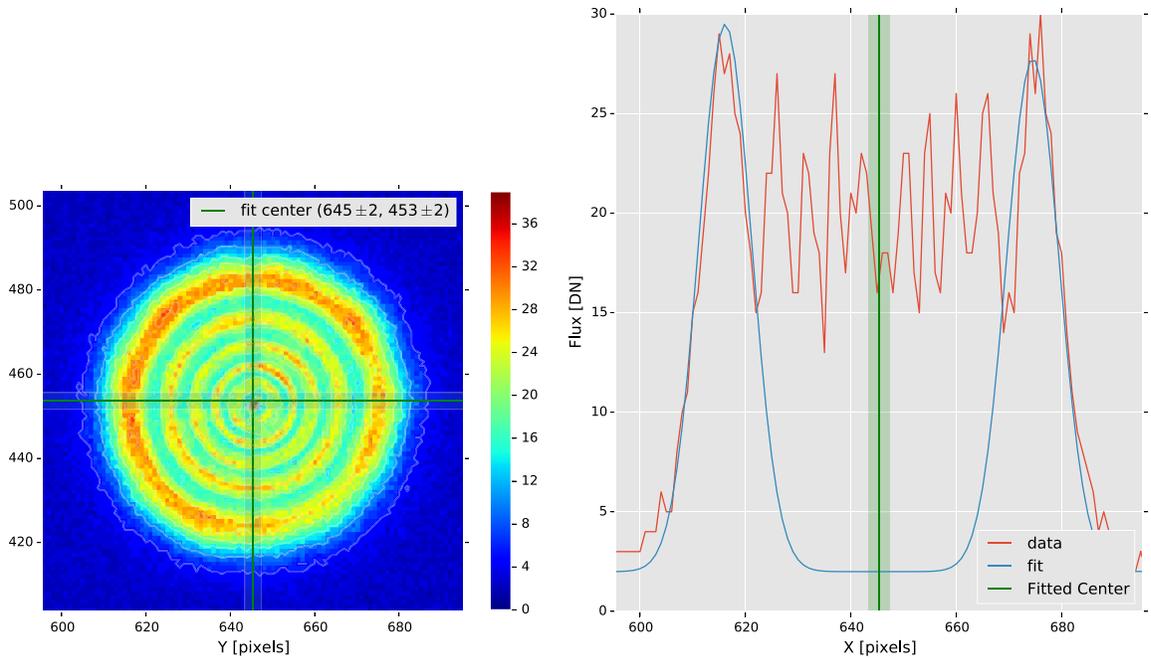

Fig. 8. (*Left*) An image of the laser spot as observed by the SAAS camera. Fresnel diffraction cleanly outline the center of the spot providing a straightfoward method to determine the calibrated center shown. The results of a Gaussian-profile ring is fit to outermost diffraction ring. The results of the fit are shown over the plot. (*Right*) A cutout through the center is shown along with the fit.

During testing it was found that the calibrated center shifted significantly when the detectors were cold compared to room temperature. Further testing was performed to characterize this effect and it was found that the shift was essentially linear as a function of temperature with a slope of 2.3 arcsec per degree Celsius. All alignments were therefore repeated with the detector cold. Since the detector temperature are maintained to within a few degrees this systematic error can be ignored.

This calibration was performed both before and after a vibration test of the payload. It was found that the calibrated center moved approximately 2 arcmin. The cause of this shift is not determined but may be due to the shifting of the Aluminum optical bench.

In order to determine the center of the X-ray detectors, an X-ray generator (Trufocus TFS-3007-HP) with a focal spot size of 0.5 mm (5 arcsec) is used to illuminate each optic in turn to generate an X-ray image. An object at this distance generates an image that is 22 cm behind the detector plane therefore the image is of concentric rings with radii ranging from 5.7 mm to 4.2 mm (3.6 mm for the 10-shell optic). The angle of the optic with respect to the source can be measured by the asymmetry of the flux distribution around the ring while the center of the rings gives the position of the source. An image of this for detectors 5 (CdTe) and detector 6 (Si) can be seen in Fig. 9.

A fit (similar to that performed on the SAAS image) provides the optical center of the detector to within about one detector pixel (∼7 arcsec). With these images combined with the SAAS image, the optical axis is determined for each detector with respect to the SAAS optical center. The next and final step is to measure the SAAS offset with respect to the SPARCS LISS.

### 3.6. *Ground solar observations*

Since it is not practical to point the payload at and track the Sun on the ground, a heliostat was used to illuminate the SPARCS and SAAS simultaneously. The heliostat system consists of a single mirror mounted to a precision motor on an equatorial mount. This solar observation provided an important measurement of the throughput of the system which informed the choice of neutral density filters for the flight as well as providing a measure of the angular resolution of the system. The SPARCS LISS is a differential detector system which uses a quad photocell and provides a difference signal between opposite cells. In order to measure the SAAS to LISS offset, the LISS signal was used to drive the heliostat mirror and zero its signal aligning the Sun with the LISS. An image of the Sun was then taken with the SAAS (see Fig. 10) and the Sun center was determined. This center was then compared with the center determined by the X-ray alignment. The offset was used to calculate the size of shims which were placed below the legs of the LISS in order to reduce the offset. Once the final set of shims were placed, a final SAAS image was taken to determine the instrument to LISS offset. The measured LISS to SAAS offset was found to be ∼ 2.9 ± 0.1 arcmin where the error is a combination of jitter in the location of the Sun and error in determining the Sun center.

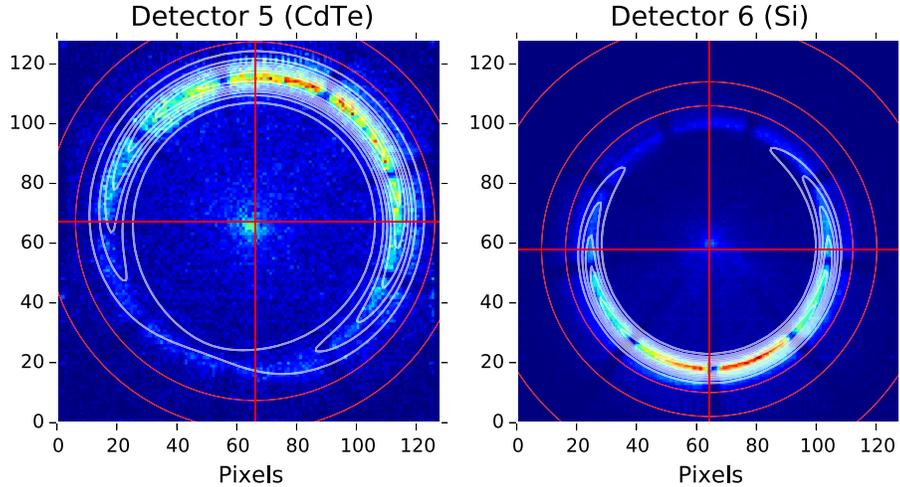

Fig. 9. An image of the X-ray source during alignment as recorded by a CdTe and Si FOXSI detector. Since the telescope is focused at infinity, this image shows a defocused image representing the rays from the optics coming to a focus behind the detector plane. The center of the ring is the focus while the asymmetry in the illumination is a measure of the tilt of the optics. The gray contours represent a fit to the data which returns the center to an accuracy of about one detector pixel (∼7 arcsec).

## 3.7. *Performance*

The resolution of the SAAS was evaluated by investigating the solar limb which is known to be extremely sharp (sub-arcsecond) in the wavelength range observed by the SAAS (Henneck *et al.*, 1999; Fivian *et al.*, 2002). To evaluate the angular resolution achieved by the SAAS, the limb of the Sun as observed by the SAAS was compared to a near-simultaneous SDO HMI (Schou *et al.*, 2012) white light image which observes at a similar wavelength (617.3 nm). The HMI image was blurred with a Gaussian filter until the limb profile matched that of the SAAS image. The Gaussian width necessary to achieve this was found to be ≈ 20 arcsec giving a rough measure of the SAAS angular

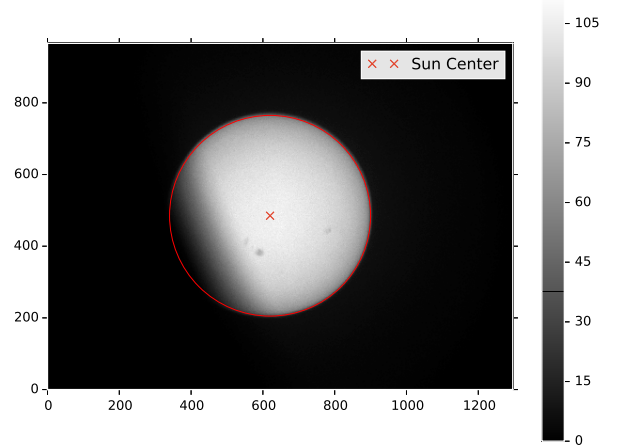

Fig. 10. (Color online) An image of the Sun taken by the SAAS during an alignment with the SPARCS system. The image of the Sun is partially occulted due to the limited size of the heliostat mirrors. Small active regions can be seen in this image and a fit to the location of the solar limb is shown as a red line. The center of this fit represents the SPARCS LISS optical center. A comparison to the X-ray alignment center shows that the LISS to SAAS offset was 2.9 ± 0.1 arcmin where the error is a combination of jitter in the location of the Sun due to winds moving the heliostat mirror combined with error in determining the Sun center.

resolution convolved with the atmospheric seeing conditions on that day.

Post-flight comparison of solar X-ray observations show that the misalignment between the X-ray optics and the LISS were much larger than measured on the ground. The exptect offset with the LISS was $\sim 3 \pm 2$ arcmin where the error term is from the shift observed during vibration testing. Flight data shows that the on-flight misalignment was 7.1 arcmin. This larger offset is likely due to the fact that the rocket motor suffered from an instability which exposed the payload to vibration loads much larger than expected or tested for.

## 4. Conclusion

This paper has described some of the major upgrades to the FOXSI for the second launch of the payload. These upgrades included adding 2 new 10 shell optic modules which increased the total effective area of FOXSI by about 20%. The detectors were also updated to include two new CdTe detectors with 60 micron pitch compared to the 75 micron pitch Si detectors. Finally a new optical telescope, the new SAAS was described which provided a new and more accurate alignment measurement between the X-ray optics and the SPARCS Sun-pointing system. Unfortunately the payloads experienced very high vibration loads which led to a much larger offset than measured by the new alignment process. An updated alignment measurement will make use of an autocollimator to both focus the SAAS and measure the alignment between the LISS and the SAAS. This new procedure should remove limitations by atmospheric seeing and the heliostat.

## Acknowledgments

This material is based upon work supported by the National Aeronautics and Space Administration under Grant NNH09ZDA001N-SHP No. issued through the Science Mission Directorate's Research Opportunities in Space and Earth Sciences (ROSES) Program. The authors would like to thank the NSROC team without which this mission would not have been a success.